\documentclass[useAMS,usenatbib]{mn2e}

\usepackage{amssymb} \usepackage{amsmath} \usepackage{times}
\usepackage{graphicx}

\newcommand{\lya}{Ly$\alpha$}

\newcommand{\msun}{\mbox{$M_\odot$}}
\newcommand{\Zsun}{\mbox{$Z_\odot$}}

\newcommand{\kms}{\mbox{km s$^{-1}$}}

\newcommand{\apjs}{ApJS}
\newcommand{\apj}{ApJ} 
\newcommand{\aj}{AJ} 
\newcommand{\araa}{ARA\&A}
\newcommand{\apjl}{ApJL} 
\newcommand{\nat}{Nature}
\newcommand{\pasp}{PASP} 
\newcommand{\aap}{A\&A}
\newcommand{\mnras}{MNRAS}

\def\ltsima{$\; \buildrel < \over \sim \;$}
\def\simlt{\lower.5ex\hbox{\ltsima}} \def\gtsima{$\; \buildrel > \over
  \sim \;$} \def\simgt{\lower.5ex\hbox{\gtsima}}

\title[Rest-Frame UV Spectrum of `The Cosmic Eye'] {A Study of 
Interstellar Gas and Stars in the Gravitationally Lensed 
Galaxy `The Cosmic Eye' from Rest-Frame 
Ultraviolet Spectroscopy}

\author[A. M. Quider et al.]{Anna M. Quider$^{1}$\thanks{Email:
    aquider@ast.cam.ac.uk},  Alice E. Shapley$^2$, Max Pettini$^1$,
  Charles C. Steidel$^3$, and
  \newauthor  Daniel P. Stark$^1$\\  
  $^1$ Institute of Astronomy, Madingley Rd, Cambridge, CB3 0HA, UK\\  
  $^2$ Department of Physics and Astronomy, University of California, 
  Los Angeles,  CA 90095-1547, USA\\  
  $^3$ California Institute of Technology, Mail Stop 105-24, Pasadena, CA
  91125, USA}

\begin{document}

\date{Accepted ... Received ... in original form ...}

\pagerange{\pageref{firstpage}--\pageref{lastpage}} \pubyear{2009}

\maketitle

\label{firstpage}

\begin{abstract}
We report the results of a study of the rest-frame ultraviolet (UV)
spectrum of the Cosmic Eye (J213512.73$-$010143), a luminous
($L \sim 2 L^\ast$)
Lyman break galaxy at $z_{\rm  sys}=3.07331$ magnified
by a factor of $\sim 25$ via gravitational lensing
by foreground mass concentrations at $z = 0.73$ and 0.33.  
The spectrum,
recorded at high resolution and signal-to-noise ratio with 
the ESI spectrograph on the Keck\,{\sc ii} telescope,
is rich in absorption 
features from the gas and massive stars in this galaxy.  
The interstellar absorption lines are resolved into two
components of approximately equal strength
and each spanning several hundred \kms\ 
in velocity. One component has a net blueshift of
$-70$ \kms\ relative to the stars and H\,{\sc ii}
regions and presumably arises in a galaxy-scale
outflow similar to those seen in most star-forming galaxies
at $z = 2$--3. The other is more unusual in showing a mean
\textit{redshift} of $+350$ \kms\  relative to $z_{\rm sys}$;
possible interpretations include
a merging
clump, or material ejected by a previous star formation episode
and now falling back onto the galaxy, or more simply a chance
alignment with a foreground galaxy. 
In the metal absorption lines,
both components  only partially cover the OB stars against
which they are being viewed. However,  there must also be more 
pervasive diffuse gas to account for the near-total covering 
fraction of the strong damped \lya\ line, indicative of  
a column density 
$N$(H\,{\sc i})\,$ = (3.0 \pm 0.8) \times 10^{21}$\,cm$^{-2}$.
We tentatively associate this neutral 
gas with the redshifted component, and propose that it 
provides the dust
`foreground screen' responsible for the low ratio
of far-infrared to UV luminosities of the Cosmic Eye.

The C\,{\sc iv} P~Cygni line in the stellar spectrum is
consistent with continuous star formation with a Salpeter initial mass
function, stellar masses from 5 to 100\,\msun, and a
metallicity $Z \sim 0.4 Z_{\odot}$.  
Compared to other well-studied examples of 
strongly lensed galaxies, we find that the young stellar 
population of the Cosmic Eye is essentially indistinguishable
from those of the Cosmic Horseshoe and MS\,1512-cB58.
On the other hand, the interstellar spectra of all three galaxies
are markedly different, attesting to the real complexity of the
interplay between starbursts and ambient interstellar matter
in young galaxies observed during the epoch when cosmic star formation
was at its peak.
\end{abstract}


\section{Introduction}
\label{sec:introduction}

High redshift star-forming galaxies are often identified by a break in
their ultraviolet continuum that is due to the Lyman limit,
partly from interstellar (within the galaxy) and primarily
from intergalactic H\,{\sc i} absorption below 912\,\AA\
(Steidel et al. 1996).  Since the discovery of these
`Lyman break galaxies' (or LBGs), samples of 
galaxies at $z = 2$--4 have increased a thousand-fold.  
Despite the variety of methods now employed to select high redshift
galaxies, LBGs remain the most common and most extensively studied
class of such objects.

Detailed studies of the physical properties of \textit{individual} 
galaxies have in general been limited by their faintness 
($m_{\cal R}^{\ast} = 24.4$, Steidel et al. 1999;
Reddy et al. 2008).  In a few cases, however,
gravitational lensing by foreground massive galaxies,
groups, or clusters has afforded rare insights into the 
stellar populations and interstellar media of 
galaxies at $z = 2$--4 
(e.g. Pettini et al. 2000, 2002; Teplitz et al. 2000; 
Lemoine-Busserolle et al. 2003; Smail et al. 2007;
Swinbank et al. 2007; 
Cabanac, Valls-Gabaud, \& Lidman 2008;
Siana et al. 2008, 2009;
Finkelstein et al. 2009; Hainline et al. 2009;
Quider et al. 2009; Swinbank et al. 2009; Yuan \& Kewley 2009
and references therein).  
The order-of-magnitude boost in flux provided
by strong lensing makes it possible to record
the spectra of these galaxies with a combination
of high resolution and high signal-to-noise ratio (S/N) which 
for normal, unlensed galaxies will have to wait until the
next generation of 30+\,m optical/infrared telescopes.

Focusing on the rest-frame UV spectra in particular,
our group has so far published observations of 
two of these strongly lensed high redshift galaxies:
MS\,1512-cB58 and J1148+1930 (cB58 and the `Cosmic Horseshoe'
respectively; Pettini et al. 2002; Quider et al. 2009).  Both are
$\sim L^{*}$ galaxies at redshifts $z\sim 2.5$
magnified by a factor of $\sim 30$ (Seitz et al. 1998;
Dye et al. 2008).  A considerable
amount of data that are inaccessible to typical low-resolution
studies is contained in these rest-UV spectra, including information
on interstellar gas composition and kinematics, on the initial mass 
function (IMF) of starbursts at high $z$, 
and clues to the geometry of the gas and stars.

One of the motivations for this work is 
to assess the range of properties possessed 
by high redshift star-forming galaxies.  
By comparing the UV spectra
of cB58 and the Cosmic Horseshoe, 
Quider et al. (2009) showed that the young 
stellar populations of these galaxies are very similar, 
as are the metallicities and kinematics of
their interstellar media.  
On the other hand, the two galaxies exhibit clear differences
in the covering fractions of their stars by the interstellar gas, 
differences which are reflected in the strikingly different morphologies
of their Lyman~$\alpha$ lines (a damped absorption profile
in cB58 and a double-peaked emission profile in the Horseshoe). 
Studies at other wavelengths (Teplitz et al. 2000; 
Hainline et al. 2009) have also shown that
cB58 and the Cosmic Horseshoe have comparable star 
formation rates, $\rm{SFR} \sim 50$--100\,\msun~yr$^{-1}$,
at the upper end of the range of values measured in 
star-forming galaxies at $z = 2$--3, 
and similar dynamical masses, 
$M_{\rm dyn} \sim 1 \times 10^{10}$\,\msun, 
typical of luminous galaxies at this epoch 
(Pettini et al. 2001; Erb et al. 2006b,c;  Reddy et al. 2006).

Clearly, several examples of strongly lensed galaxies 
need to be studied at such levels of detail for a full 
characterization of galaxy properties at these redshifts.
Fortunately, many new cases have been discovered
recently, mostly from dedicated searches in imaging data
from the Sloan Digital Sky Survey (Estrada et al. 2007; 
Belokurov et al. 2007, 2009; Ofek et al. 2008; Shin et al. 2008;
Kubo et al. 2009;
Lin et al. 2009; Wen et al. 2009).
Other surveys have capitalised on the superior
spatial resolution of 
\textit{Hubble Space Telescope (HST)} to identify
high redshift galaxies strongly lensed into multiple
images or arcs; the $z = 3.0733$ galaxy which is the focus of the
present paper, dubbed the `Cosmic Eye', was indeed
discovered by Smail et al. (2007) from an \textit{HST}
Snapshot program targeting high luminosity X-ray clusters.
  
In this paper we present high resolution observations
of the rest-frame UV spectrum of the Cosmic Eye.  
The paper is organized as follows.  
In Section~\ref{sec:eyesummary} we summarize 
the known properties of the Cosmic Eye from
previous studies at a variety of wavelengths, while
Section~\ref{sec:observations} has details of our observations
and data reduction.  
We analyze the interstellar spectrum of the 
Cosmic Eye in Section~\ref{sec:ISM}, and the composite
spectrum of its young stellar population in 
Section~\ref{sec:stars}.
Section~\ref{sec:intervening}
focuses on the numerous intervening absorption systems that
potentially confound our interpretation of the galaxy's intrinsic
features.  In Section~\ref{sec:discussion}
we discuss the implications of our findings;
finally, we summarize our main conclusions in 
Section~\ref{sec:conclusions}.
Throughout the
paper, we adopt a cosmology with $\Omega_{\rm M} = 0.3$,
$\Omega_{\Lambda} = 0.7$, and $H_0 = 70$\,km~s$^{-1}$~Mpc$^{-1}$.

\section{The Cosmic Eye}
\label{sec:eyesummary}

\begin{figure}
\vspace*{0.25cm} 
 \centerline{\hspace{-0.15cm}
  \includegraphics[width=0.95\columnwidth,clip,angle=0]{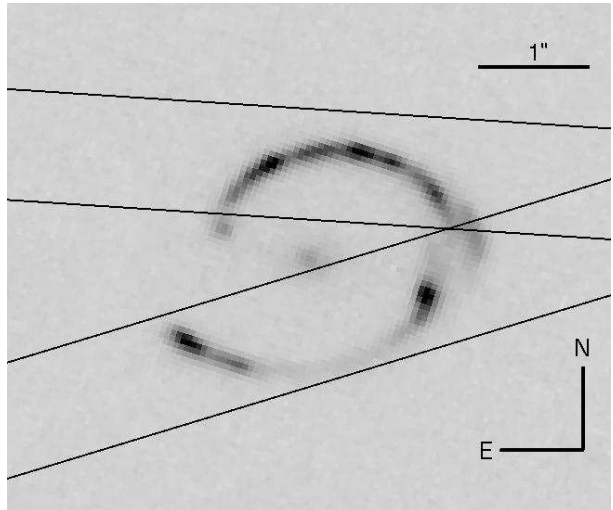}}
\caption{HST/ACS F606W image of the Cosmic Eye.  
The straight lines indicate the two placements of the
$1.0 \times 20^{\prime\prime}$ entrance slit of ESI 
used for the observations reported
here.  }
\label{fig:eyeimage}
\end{figure}

Among the bright, strongly lensed, high redshift galaxies,
the Cosmic Eye is the most well-studied to date after
cB58. Here we summarize
available information which is relevant to our analysis.  

\begin{figure*}
\vspace{-3.75cm} 
\centerline{\hspace{-0.15cm}
  \includegraphics[width=1.725\columnwidth,clip,angle=270]{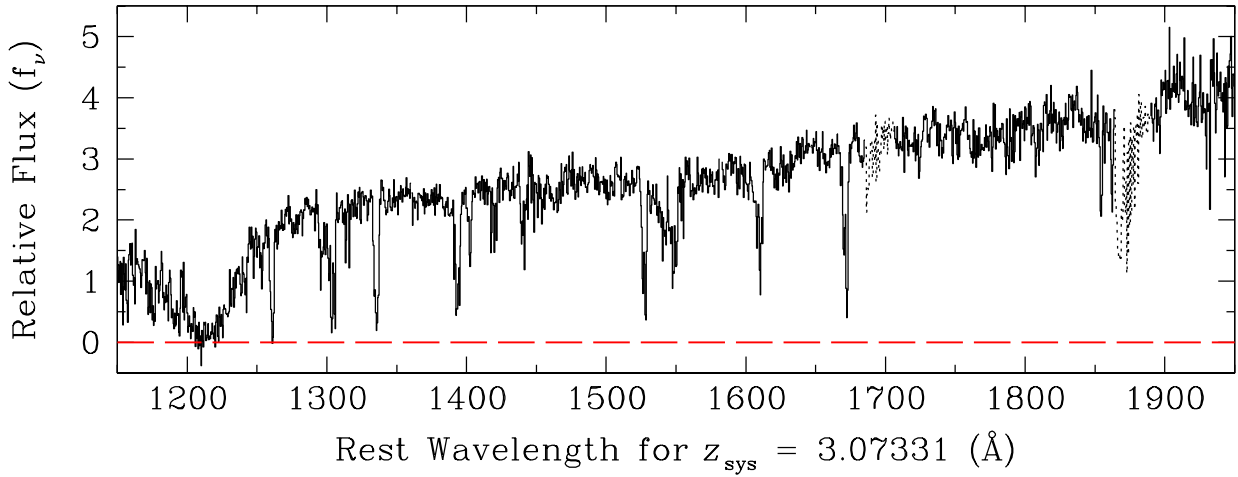}}
\vspace{-4.00cm}
\caption{The UV spectrum of the Cosmic Eye between 
rest-frame wavelengths 1150\,\AA\ and 1950\,\AA. The flux 
from both northern an southern arcs was summed to produce
this composite, which was rebinned to 2\,\AA\ bins (to facilitate
reproduction) before reducing to the rest-frame at $z_{\rm sys} = 3.07331$.
Regions of the spectrum affected by atmospheric absorption are plotted 
with a dotted line.}
\medskip
\label{fig:fullspectrum}
\end{figure*}

The Cosmic Eye, also known as LBG J213512.73$-$010143, was 
so named because it consists  of two
bright arcs extending $\sim 3^{\prime\prime}$ which nearly fully surround a
compact galaxy (Figure \ref{fig:eyeimage}).  From low resolution
spectroscopy, the central galaxy was classified as a massive early-type
spiral, or S0, galaxy at $z = 0.73$, and from rest-frame optical spectroscopy the
source redshift was estimated to be $z=3.0743$ (Smail et
al. 2007).  The detailed lensing model by Dye et al. (2007) concluded
that the lensing of the Cosmic Eye is attributable to both the galaxy
at $z=0.73$ and a foreground cluster, MACS J2135.2-0102, at
$z=0.33$. Their models require two source components at similar
redshifts: a main component which creates the Eye's ringed structure,
and a secondary component, offset $\sim 0.3^{\prime\prime}$ to the
west, responsible for producing faint extensions
at either end of the northern arc. 
The total magnification of the system is estimated to be $\sim 25$;
when this is taken into account, the observed magnitude $r=20.3$ 
corresponds to an intrinsic luminosity 
$L_{\textrm 1700}\sim 1.5$--$2 L^{\ast}_{\textrm 1700}$
compared to the luminosity function of $z = 3$ galaxies
estimated by Steidel et al. (1999) and more recently Reddy et al. (2008).

The geometry and kinematics of the Cosmic Eye have begun to be explored.
Most recently, Siana et al. (2009) used the galaxy's far-infrared luminosity,
which they found to be $\sim 8$ times lower than expected from its
rest-frame UV luminosity, to argue that a reddening curve similar
to those determined from observations of stars in
the Large and Small Magellanic Clouds (LMC/SMC) is
more appropriate for the dust in the Cosmic Eye than the Calzetti et al.
(2000) attenuation curve normally adopted for starbursts.  
The difference between the two reddening laws is one
of geometry: the Calzetti et al. (2000) curve is thought to arise
from a configuration where stars and dust are mixed together,
whereas the LMC/SMC (or. for that matter, Milky Way)
extinction curves are more appropriate to the reddening 
produced by a `screen of dust' which is in the foreground
of the sources of UV light.

Integral field spectroscopy
of H$\beta$ and [O\,{\sc iii}]\,$\lambda \lambda 4959,
5007$ nebular emission lines with the
OH-Suppressing Infrared Integral Field Spectrograph
(OSIRIS) on the Keck~{\sc ii} telescope
(Stark et al. 2008) showed a regular rotation pattern
$v_{\rm rot} \sin i= 55$\,\kms\ on which are superimposed
random motions with a comparable 
velocity dispersion of $\sigma= 54$\,\kms.
Coppin et al. (2007) detected  
CO  emission  from the Cosmic Eye and found the
signal to peak at the location of the secondary
component of the source, suggesting that this
component may be the 
remaining reservoir of gas available to fuel 
star formation.

The star formation rate of the galaxy has been estimated
from its luminosity in the UV stellar continuum, 
H$\beta$ emission line, and infrared dust emission to be
${\rm SFR} \sim 100$, $\sim 100$, and $\sim 60-140$ \msun\,yr$^{-1}$,
respectively (Smail et al. 2007; Stark et al. 2008; Coppin et
al. 2007; Siana et al. 2009).\footnote{All of these estimates
are based on the conversion factors between luminosity and SFR
given by Kennicutt (1998) and are appropriate for a Salpeter (1955) IMF.
Adoption of a more realistic formulation of the IMF, 
such as that proposed by Chabrier (2003), would result
in smaller values of SFR by a factor of 1.8\,.}
The metallicity determined from the ratio of strong
emission lines (the $R_{23}$ method of Pagel et al. 1979), 
is $Z \sim 0.9$\,\Zsun\ (Stark et al. 2008).  
At the current rate of star formation,
it would have taken the galaxy 
$\sim 100$\,Myr  to build its
stellar mass of $\sim 6\times 10^{9}$\,\msun\
(Coppin et al. 2007).

To this growing body of data we now add a high resolution
study of the rest-frame UV spectrum of the Cosmic Eye.
At UV wavelengths the spectrum is dominated by the
combined emission from early-type stars and absorption
from interstellar gas. Such data can shed new light on the 
kinematics, chemical composition and stellar populations
of this galaxy, complementing the information gleaned
from other wavelengths.

\begin{figure*}
\vspace{-3.85cm} 
\centerline{\hspace{0.15cm}
  \includegraphics[width=1.5\columnwidth,clip,angle=270]{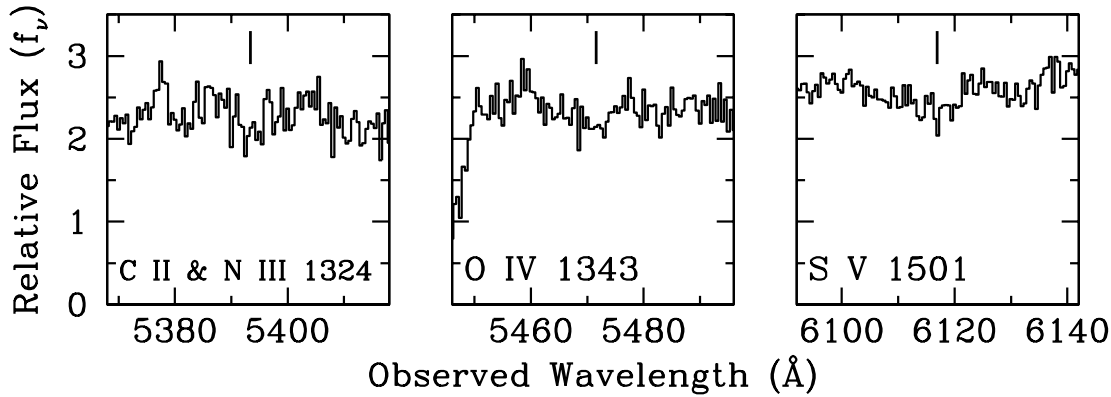}}
\vspace{-3.15cm}
\caption{
Portions of the spectrum of the Cosmic Eye encompassing stellar photospheric
absorption lines used for the determination of the systemic redshift of the galaxy
(see Table~\ref{tab:systemic_z}).
}
\medskip
\label{fig:stellarlines}
\end{figure*}

\section{Observations and Data Reduction}
\label{sec:observations}

The procedures we followed for our observations and
data reduction were very similar to those described by
Quider et al. (2009) in their study of the Cosmic Horseshoe,
and we refer the interested reader to that paper for a 
comprehensive account. 
Briefly, we used
the Echellette Spectrograph and Imager (ESI; Sheinis et
al. 2000) on the Keck\,{\sc ii} telescope 
to record the rest-frame UV spectrum of the Cosmic Eye
on the nights of 2007 September 13 and 14 UT.  
With its high efficiency, wide wavelength coverage
(from $\sim 4000$ to 10\,000\,\AA),
and moderately high resolution 
($R \equiv \lambda/\Delta \lambda = 4000$),
ESI is the instrument of choice for this work.
On the first night, the $1^{\prime\prime}$ wide slit of ESI 
was positioned on the northern arc of the Cosmic Eye
at position angle $\textrm{P.A.} = 86.44^\circ$ 
(see Figure~\ref{fig:eyeimage}); on the second night
we recorded the light of the southern arc at
$\textrm{P.A.} = 287.12^{\circ}$.
The total exposure time \textit{for each arc}
was 16\,000\,s, broken up into eight 2000\,s long integrations
on the ESI detector. The seeing was sub-arcsecond on both nights.

Inspection of the two-dimensional (2D) 
spectra shows that only the light from the main 
ring structure was recorded; the fainter extensions to the east and the west
of the northern
ring which, as explained above (Section~\ref{sec:eyesummary}),
are due to the secondary component of the source
in the lensing model of Dye et al. (2007), were only 
partially covered by the slit placements (see Figure~\ref{fig:eyeimage})
and are in any case much fainter than the main ring structures.

The data were processed with standard \textsc{iraf} tasks 
following the steps outlined in Quider et al. (2009);
compared to that work, subtraction of
the background was considerably
more straightforward because in the present case
light from the lensing galaxy did not fall within the ESI slit
(see Figure~\ref{fig:eyeimage}).
Nevertheless, once the data were fully reduced as described below,
we found residual flux in the core of the damped \lya\ absorption
line (see Section~\ref{sec:Lya}), 
amounting to $\sim 5\%$ of the continuum level.
We cannot determine whether the effect is real, indicating that
the damped \lya\ line only covers 95\% of the OB stars producing
the continuum near 1216\,\AA, or is an artifact of errors in the
determination of the background level to be subtracted.
Nor do we have any other `markers'  in the spectrum
as definitive as the damped \lya\ line to help us assess
whether this background correction, whatever its origin, 
is constant along the spectrum or varies with wavelength.
In the circumstances, we limited ourselves to applying 
to the final one-dimensional (1D) spectrum 
a uniform background correction of 5\% of the continuum flux.

At the S/N ratio of the present data, we could not find any 
significant differences between the extracted 1D
spectra of the northern and southern arcs;  
their similarity is consistent with the lensing
model of Dye et al. (2007) which shows them to be images of 
the same source, mostly contained within the caustic
(see Figure 4 of Dye et al. 2007).\footnote{Note, however,
that Smail et al. (2007) did report differences in the profiles
of the interstellar absorption lines between the northern and
southern arcs. The fact that we do not confirm such differences 
could be due either to the higher S/N ratio and resolution of our
spectra, or to the different slit positions employed in the two studies.}
Therefore, we averaged the spectra of the two
arcs and rebinned the resulting spectrum to 0.5\,\AA\ bins
to maximise the S/N ratio.
The final composite 1D spectrum (reproduced in  Figure~\ref{fig:fullspectrum})
has an average
${\rm S/N} \simeq  14$ per 0.5\,\AA\ bin
between  5200 and 7500\,\AA\ (1275--1840\,\AA\ in the rest-frame
of the source), and a resolution ${\rm FWHM} = 75$\,\kms, sampled
with $\sim 3$ wavelength bins.

\begin{table}
\centering {\hspace*{0.75cm}\begin{minipage}[c]{1.0\textwidth}
 \caption{\textsc{Systemic Redshift}}
   \begin{tabular}{llll} 
\hline  \hline Ion & $\lambda_{\rm lab}^{\rm a}$ (\AA) & $z$ & Origin\\  \hline  
C\,{\sc ii}+N\,{\sc iii} & $1324.1418^{\rm b}$ & 3.0731 & Stars\\ 

O\,{\sc iv} & 1343.354 & 3.0730 & Stars\\ 

S\,{\sc v}   & 1501.763  & 3.0732   &   Stars\\ 

H$\beta^{\rm c}$ & 4862.721 & 3.0737&  H\,{\sc ii} Regions\\

[O\,{\sc iii}]$^{\rm c}$ & 4960.295 & 3.0734 &  H\,{\sc ii} Regions\\

[O\,{\sc iii}]$^{\rm c}$ & 5008.239 & 3.0735 &  H\,{\sc ii} Regions\\

\hline
     \label{tab:systemic_z}
 \end{tabular}
 \smallskip

 $^{\rm a}$ Vacuum wavelengths.\\ 
 $^{\rm b}$ Central wavelength of the blend.\\ 
 $^{\rm c}$ Measured from OSIRIS data reported by Stark et al. (2008).
  \end{minipage}
 }
 \end{table}

\section{The Interstellar Spectrum}
\label{sec:ISM}

\subsection{The systemic redshift of the Cosmic Eye}
\label{sec:z_sys}

\begin{figure*}
\centerline{\hspace{-0.15cm}
  \includegraphics[width=1.525\columnwidth,clip,angle=270]{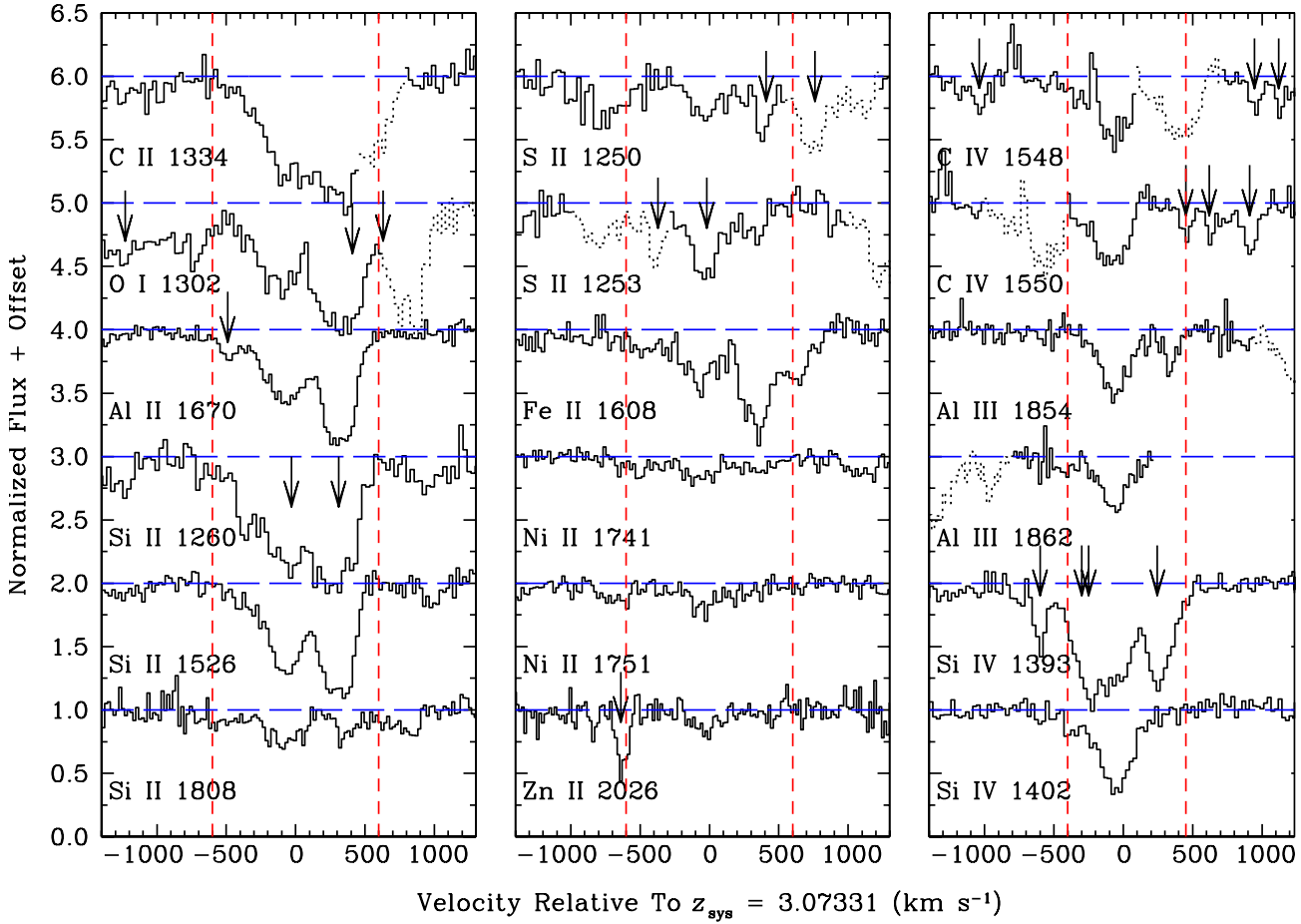}}
\caption{Normalized profiles of selected interstellar absorption
lines.  \emph{Left and middle panels}: Transitions of ions which are dominant in H\,{\sc i}
regions.  \emph{Right panel}: Transitions of ions which are ionized 
beyond the species dominant in H\,{\sc i}
regions. The vertical red dash lines indicate the velocity range
over which we measured the equivalent widths of the strongest
absorption lines (see Table~\ref{table:is_lines}). Arrows mark the
locations of intervening absorption lines unrelated to the Cosmic Eye;
these are sometimes resolved as 
individual features (e.g. see panel for Zn\,{\sc ii}\,$\lambda 2026$),
while in other cases they add to absorption from the Cosmic Eye
(e.g. see panel for Si\,{\sc ii}\,$\lambda 1260$).
Section~\ref{sec:intervening} and Table~\ref{tab:Intervening} give a full
account of these intervening absorption systems.
Dotted lines have been used to indicate spectral features 
in the Cosmic Eye other than those to which the label in each plot
refers. Thus, for example, C\,{\sc iv}\,$\lambda 1550$ is shown 
with a dotted line in the  C\,{\sc iv}\,$\lambda 1548$ plot (top plot
in the right-hand panel), and C\,{\sc iv}\,$\lambda 1548$ is shown
with a dotted line in the C\,{\sc iv}\,$\lambda 1550$ plot
(second plot from the top of right-hand panel).
}
\medskip
\label{fig:ISMvelocity3}
\end{figure*}

An accurate determination of the systemic redshift of the lensed
galaxy in the Cosmic Eye is
critical to the rest of our analysis.  
Our rest-frame UV spectrum
includes several absorption
features from the photospheres of hot stars 
which can be used for this purpose. 
The cleanest among these are 
O\,{\sc iv}\,$\lambda 1343.354$, S\,{\sc v}\,$\lambda 1501.763$,
and a close blend of  C\,{\sc ii} and\ N\,{\sc iii} lines
centred at $ \lambda  1324.1418$
(all vacuum wavelengths); see 
Figure~\ref{fig:stellarlines}.
Table \ref{tab:systemic_z} lists the redshifts deduced from
each of these stellar lines, as well as the values of redshift 
we measured from the
H$\beta$ and [O\,{\sc iii}]\,$\lambda \lambda 4959,5007$ emission
lines in the OSIRIS spectrum of Stark et al. (2008).
As can be seen from Table~\ref{tab:systemic_z},
the mean redshift of the UV stellar lines,
$\langle z_\textrm{OB stars}\rangle = 3.0731$,
differs by $-29$\,\kms\ from the mean
$\langle z_{\rm H\,\textsc{ii}}\rangle = 3.0735$
of the three nebular lines.\footnote{Note that this differs
from $\langle z_{\rm H\,\textsc{ii}}\rangle = 3.0743$
reported by Smail et al. (2007), but the latter was
not on a vacuum heliocentric scale (M. Swinbank, private communication).}
This small difference may reflect a small offset between the
wavelength calibrations of ESI and OSIRIS, or may
be simply due to noise in the spectra. It could also be real
but, as we have no other reason to suspect that the massive
stars in this galaxy and the nebulae they ionize should
be at systematically different redshifts, we decided to average
together all the values of redshift in Table~\ref{tab:systemic_z}
to deduce a mean systemic redshift for the galaxy
$z_{\rm sys} = 3.07331 \pm 0.00024$ ($1 \sigma$).
The standard deviation corresponds to a velocity
uncertainty $\delta v = 18$\,\kms, or just over one half
of a wavelength bin in the final 1D ESI spectrum.

\subsection{Kinematics of the absorbing gas}
\label{sec:kinematics}

\begin{table*}
\centering
 \begin{minipage}[c]{1.0\textwidth}
 \caption{\textsc{Interstellar Absorption Lines}}
   \begin{tabular}{@{}llllccclcccl} 
\hline 
\hline
     \multicolumn{1}{c}{Ion}
&  \multicolumn{1}{c}{$\lambda_{\rm lab}^{\rm a}$}  
&  \multicolumn{1}{c}{$f^{\rm a}$} 
&  \multicolumn{1}{c}{$\Delta v_{\rm blue}^{\rm b}$} 
&  \multicolumn{1}{c}{$z_{\rm blue}^{\rm c}$} 
&  \multicolumn{1}{c}{$W_{\rm blue}^{\rm d}$} 
&  \multicolumn{1}{c}{$\delta W_{\rm blue}^{\rm e}$} 
&  \multicolumn{1}{c}{$\Delta v_{\rm red}^{\rm f}$} 
&  \multicolumn{1}{c}{$z_{\rm red}^{\rm g}$} 
&  \multicolumn{1}{c}{$W_{\rm red}^{\rm h}$} 
&  \multicolumn{1}{c}{$\delta W_{\rm red}^{\rm i}$} 
&  \multicolumn{1}{c}{Comments}\\
   \multicolumn{1}{c}{ }
& \multicolumn{1}{c}{(\AA)}
& \multicolumn{1}{c}{ }
& \multicolumn{1}{c}{(km~s$^{-1}$)}
& \multicolumn{1}{c}{ }
& \multicolumn{1}{c}{(\AA)}
& \multicolumn{1}{c}{(\AA)}
& \multicolumn{1}{c}{(km~s$^{-1}$)}
& \multicolumn{1}{c}{ }
& \multicolumn{1}{c}{(\AA)}
& \multicolumn{1}{c}{(\AA)}
& \multicolumn{1}{c}{ }\\
   \multicolumn{1}{c}{(1) }
& \multicolumn{1}{c}{(2)}
& \multicolumn{1}{c}{(3) }
& \multicolumn{1}{c}{(4)}
& \multicolumn{1}{c}{(5) }
& \multicolumn{1}{c}{(6)}
& \multicolumn{1}{c}{(7)}
& \multicolumn{1}{c}{(8)}
& \multicolumn{1}{c}{(9) }
& \multicolumn{1}{c}{(10)}
& \multicolumn{1}{c}{(11)}
& \multicolumn{1}{c}{(12) }\\
\hline
C\,{\sc ii}   & 1334.5323	& 0.1278    & $-600,+600^{\rm j}$   & 3.0752$^{\rm j}$   & 3.15$^{\rm j}$  & 0.06$^{\rm j}$   & \ldots           & \ldots  & \ldots  & \ldots    & Blended with C\,{\sc ii}$^{\ast}\,\lambda 1335.6627$\\
C\,{\sc iv}  & 1548.204	& 0.1899    & $-400,+100$                & 3.0719                   & 0.68                   & 0.06                   & \ldots           & \ldots  & \ldots  & \ldots    & Blended with stellar C\,{\sc iv}\,$\lambda 1549.1$\\
	           & 1550.781	& 0.09475  & $-400,+100$                & 3.0718                   & 0.74                   & 0.04                   & \ldots           & \ldots  & \ldots  & \ldots    & Blended with stellar C\,{\sc iv}\,$\lambda 1549.1$\\
O\,{\sc i}    & 1302.1685	& 0.04887  & $-500,+75$                  & 3.0711                   & 1.12                   & 0.04                   & \ldots           & \ldots  & \ldots  & \ldots    & Redshifted component blended \\
                   &                     &              &                                    &                              &                         &                           &                     &              &             &               & with Si\,{\sc ii}\,$\lambda 1304.3702$\\
Al\,{\sc ii}  & 1670.7886	& 1.74       & $-400,+150$                 & 3.0722                   & 1.19                   & 0.02                   & $+150,+600$  & 3.0777  & 1.40      & 0.02       & \\	
Al\,{\sc iii} & 1854.7184	& 0.559     & $-400,+200$                 & 3.0725                    & 0.84                  & 0.02                   & $+200,+450$  & 3.0779  & 0.23      & 0.01       & \\
	           & 1862.7910	& 0.278	& $-400,+180$                 & 3.0722                   & 0.68                   & 0.02                   & \ldots           & \ldots  & \ldots  & \ldots   &  \\
Si\,{\sc ii}   & 1260.4221	& 1.18       & $-600, +600^{\rm j}$   & 3.0740$^{\rm j}$  & 3.20$^{\rm j}$   & 0.05$^{\rm j}$   & \ldots            & \ldots  & \ldots  & \ldots   & Blended with S\,{\sc ii}\,$\lambda 1259.519$\\
	           & 1304.3702	& 0.0863   & \ldots                          & \ldots                   & \ldots              & \ldots               &   $+85,+500$   & 3.0772  & 1.20      & 0.03       & Blueshifted component blended \\
                   &                     &              &                                    &                              &                         &                           &                      &             &             &               & with O\,{\sc i}\,$\lambda 1302.1685$\\
	           & 1526.7070	& 0.133	& $-600,+100$                 & 3.0716                   & 1.22                  & 0.03                    & $+100,+600$  & 3.0772  & 1.30      & 0.02        & \\
	           & 1808.0129	& 0.00208  & $-300,+150$                 & 3.0723                   & 0.48                  & 0.03                    & $+150,+500$  & 3.0782  & 0.24      & 0.03       & \\
Si\,{\sc iv}  & 1393.7602	& 0.513     & $-500,+120$                 & 3.0711                   & 1.83                   & 0.03                   & \ldots           & \ldots  & \ldots  & \ldots    &  \\
	           & 1402.7729	& 0.254	& $-400,+200$                 & 3.0722                   & 0.94                  & 0.03                    & \ldots           & \ldots  & \ldots  & \ldots    &  \\
S\,{\sc ii}    & 1250.578      & 0.00543  & $-300,+150$                 & 3.0727                  & 0.45                   & 0.03                    & $+150,+500$  & 3.0780  & 0.40      & 0.03       & \\
                   & 1253.805      & 0.0109    & $-300,+150$                 & 3.0726                  & 0.69                   & 0.03                    & $+150,+500$  & 3.0774  & 0.28      & 0.03       & \\
Fe\,{\sc ii}  & 1608.4511	& 0.0577    & $-300,+150$                 & 3.0724                  & 0.69                   & 0.03                   & $+150,+500$   & 3.0779  & 1.03      & 0.03      & \\
Ni\,{\sc ii}  & 1709.6042    & 0.0324    & $-300,+150$                 & 3.0727                  & 0.17                   & 0.02                   & $+150,+500$  & 3.0775  & 0.15       & 0.02      & \\
          	   & 1741.5531	& 0.0427    & $-300,+150$                 & 3.0724                  & 0.31                   & 0.03                   & $+150,+500$  & 3.0777  & 0.18       & 0.02      & \\
                    & 1751.9157	& 0.0277    & $-300,+150$                 & 3.0726                  & 0.31                   & 0.03                   & $+150,+500$  & 3.0769  & 0.12       & 0.02      & \\
Zn\,{\sc ii}  & 2026.137      & 0.501     & $-300,+150$                 & 3.0729                  & 0.27                   & 0.03                   & $+150,+500$  & 3.0783  & 0.12        & 0.02     & \\
\hline
\label{table:is_lines}
      \end{tabular}
      \smallskip
 
 $^{\rm a}$ Vacuum wavelength and $f$-values are from Morton (2003)  with  updates by Jenkins \& Tripp (2006).\\
 $^{\rm b}$ Velocity range for measurements of blueshifted absorption component.\\
 $^{\rm c}$ Redshift of the centroid of blueshifted absorption component.\\
 $^{\rm d}$ Rest-frame equivalent width of blueshifted absorption component.\\
 $^{\rm e}$ $1 \sigma$ random error on the rest-frame equivalent width of blueshifted absorption component.\\
 $^{\rm f}$ Velocity range for measurements of redshifted absorption component.\\
 $^{\rm g}$ Redshift of the centroid of redshifted absorption component.\\
 $^{\rm h}$ Rest-frame equivalent width of redshifted absorption component.\\
 $^{\rm i}$ $1 \sigma$ random error on the rest-frame equivalent width of redshifted absorption component.\\
 $^{\rm j}$ These values refer to the blend of blueshifted and redshifted absorption components.\\
\end{minipage}
 \end{table*}


We closely inspected the spectrum of the Cosmic Eye for 
interstellar absorption lines and identified 20 transitions
from eight elements in different ionization states,
from O\,{\sc i} to C\,{\sc iv}.  The absorption lines are listed in
Table~\ref{table:is_lines} and all but two (for clarity 
purposes) are reproduced in the montage in 
Figure~\ref{fig:ISMvelocity3}.\footnote{The 
two transitions omitted from Figure~\ref{fig:ISMvelocity3}
are Si\,{\sc ii}\,$\lambda 1304$, which is actually visible in the 
plot for the nearby O\,{\sc i}\,$\lambda 1302$ where it is
indicated by a dotted line, and Ni\,{\sc ii}\,$\lambda 1709$
which is of similar strength to the other two Ni\,{\sc ii} lines
reproduced in the middle panel of Figure~\ref{fig:ISMvelocity3}.}

The most striking aspect of the Cosmic Eye's interstellar absorption lines 
is that they consist of two distinct components:
a (mostly) blueshifted component centred 
at $ v_{\rm blue}  \simeq -70$\,\kms\
relative to $z_{\rm sys} = 3.07331$, and a redshifted
component centred at $ v_{\rm red}  \simeq +350$\,\kms\
(these velocities refer to the wavelength bins with the highest
apparent optical depth in the line profiles). 

As we shall see below (Section~\ref{sec:intervening}), 
the sightline to the Cosmic Eye
intersects several lower redshift absorbers
that produce a multitude of intervening absorption lines
unrelated to the lensed galaxy at $z = 3.07331$.
These features do complicate the interpretation of the
interstellar absorption due to the Cosmic Eye itself;
in Figure~\ref{fig:ISMvelocity3} we have indicated 
with downward pointing arrows instances where the
absorption lines due to the interstellar medium
in the Cosmic Eye are 
contaminated by intervening absorbers at lower redshifts.
The two-component structure of the interstellar lines
in the Cosmic Eye is most clearly seen in Figure~\ref{fig:ISMvelocity3}
in unblended transitions recorded at high S/N,
such as Si\,{\sc ii}\,$\lambda 1526.7070$ and Al\,{\sc ii}\,$\lambda 1670.7886$.
From these absorption lines it appears that the blueshifted component
extends over a velocity range $\Delta v  \sim 700$\,\kms, from 
$\sim  -600$\,\kms\ to $\sim +100$\,\kms\ relative to the systemic
redshift $z_{\rm sys} = 3.07331$, while the redshifted
component is somewhat narrower with $\Delta v  \sim 500$\,\kms\
(from $\sim +100$\,\kms\ to $\sim +600$\,\kms\ relative to $z_{\rm sys}$).
Furthermore, it can be seen from Figure~\ref{fig:ISMvelocity3}
that the redshifted component is absent in the absorption lines of the most
highly ionized gas, C\,{\sc iv} and Si\,{\sc iv}, and is relatively
weak in the intermediate ionization stage Al\,{\sc iii}.
Whenever possible, we have listed measurements of absorption
redshift and equivalent width separately for the two components
in Table~\ref{table:is_lines}. Note, however, that the values of 
equivalent width given in the table 
may include blends with intervening absorption
features, as explained above; in this respect, the purely 
random errors quoted in columns (7) and (11) underestimate
the true uncertainties (random and systematic) affecting
these measurements.

The kinematic pattern  revealed by Figure~\ref{fig:ISMvelocity3} is
unexpected for star-forming galaxies at all redshifts, where the interstellar
medium seen in absorption against the starburst is primarily moving
with \emph{negative} velocities, commonly interpreted as large-scale outflows.
For example, in the stack of more than 800 low resolution spectra 
of galaxies at $z \simeq 3$ constructed by Shapley et al. (2003), the 
interstellar lines have a net blueshift of $\sim -150$\,\kms, and
an analogous result is found 
in stacked spectra of galaxies at $z = 2$ (Law et al. 2007), and
$z = 4$, 5 and 6 (Vanzella et al. 2009).
Similarly, in her ESI study of the interstellar Na\,{\sc i} D lines
in 18 local ultraluminous infrared galaxies, Martin (2005) found 
signatures of outflows in 15 cases and of inflow in only one case.
Our earlier high-resolution work on other strongly lensed galaxies
at $z = 2 - 3$ (Pettini et al. 2002; Quider et al. 2009) did show a tail of 
absorption extending to positive velocities, but what we see in the Cosmic Eye
is qualitatively different, with the redshifted component being
of comparable strength to the blueshifted absorption.
When observed at the low resolutions of most spectra
of high-$z$ galaxies ($R < 1000$), the stronger interstellar lines
in the Cosmic Eye would have a net redshift of $\sim +150$\,km~s$^{-1}$,
which is highly unusual (e.g. Steidel  et al. in preparation).
We discuss possible interpretations in Section~\ref{sec:discuss_kinematics}.

\subsection{Partial Coverage}
\label{sec:partial_cover}

The profiles of the interstellar lines in the Cosmic
Eye hold clues to the geometry of gas, dust, and stars
in this galaxy.
An inspection of Figure~\ref{fig:ISMvelocity3} reveals that
the cores of the strongest lines which are not contaminated
by intervening absorption at lower redshifts 
(that is, lines not marked by a downward pointing arrow)
do not reach zero flux 
in either of the two kinematic components. 
We consider the implications of this residual intensity in the
line cores for each component in turn.

First we consider the redshifted component.  
Si\,{\sc ii}\,$\lambda 1526$,
Al\,{\sc ii}\,$\lambda 1670$
and Fe\,{\sc ii}\,$\lambda 1608$ 
all exhibit flat cores at a residual intensity
of $I_\lambda/I_0 \simeq 0.15$ (where $I_\lambda$ is the 
measured intensity in the line at wavelength $\lambda$ and 
$I_0$ is intensity in the continuum). 
The most straightforward interpretation
is that we are seeing the superposition within the spectrum
of saturated absorption lines and escaping continuum photons.  
Such a scenario would apply if the absorbing material only
covers $\sim 85$\% of the continuum source. 

The analysis of the blueshifted component is less clear-cut.
It appears to be covering even less
of the stellar light than the redshifted component, 
with the saturated cores of Si\,{\sc ii}\,$\lambda 1526$
and O\,{\sc i}\,$\lambda 1302$ levelling at a residual
intensity  $I_\lambda/I_0 \simeq 0.3$,  
corresponding to an apparent optical 
depth $\tau_{\rm a} \sim 1$ [with the usual
definition of the optical depth $\tau$ given by $I_\lambda/I_0 = \exp (- \tau)$]. 
Al\,{\sc ii}\,$\lambda 1670$ and Fe\,{\sc ii}\,$\lambda 1608$ 
may not be fully saturated in this component, since
their minimum residual intensities (in the blue component)
are higher: $I_\lambda/I_0 \simeq 0.45$, and $\simeq 0.55$ 
respectively. On the other hand,
we do not have a satisfactory explanation for the
higher apparent optical depth of C\,{\sc ii}\,$\lambda 1334$,
unless this transition is blended with other unrecognized
absorption (in addition to the fine structure line 
C\,{\sc ii}$^{\ast}\,\lambda 1335$ which would fall
$\sim 250$\,\kms\ to the red in the top left-hand panel of
Figure~\ref{fig:ISMvelocity3}).

In any case, a lower covering factor of the blueshifted gas
compared to the redshifted absorption is required in order to make sense of the 
apparent contradiction between the relative strengths of the red
and blue components in absorption lines of differing $f$-values.
For example, the two components are observed to have
comparable values of  $\tau_{\rm a}$ in Si\,{\sc ii}\,$\lambda 1808$ 
but different values  in 
Si\,{\sc ii}\,$\lambda 1526$, with the blue component
being weaker than the red one
(see lower two panels in the left-hand column of 
Figure~\ref{fig:ISMvelocity3}). 
Such an arrangement would at first sight appear to
be unphysical.
These two Si\,{\sc ii} transitions 
have widely different $f$-values (see Table~\ref{table:is_lines});
since they both arise from the same ground state of Si$^+$,
their \emph{intrinsic} optical depths must be in the same
ratio as their $\lambda \times f$ products, or
$\tau(1526)/\tau(1808) = 54$. This expectation can only
be reconciled with the measured values 
of \emph{apparent} optical depth, $\tau_{\rm a}$,
if we assume that the covering factor of the blue component
is lower than that of the red one, and that both components
are in fact saturated in Si\,{\sc ii}\,$\lambda 1526$.
Once that is taken into account, and the zero level adjusted accordingly,
the intrinsic (rather than
apparent) optical depth is actually higher in the blue component
than in the red one.

The combination of: (a) partial coverage by differing factors for
the blue and red components, and possibly even varying as a function
of velocity within each component (see Martin \& Bouch\'{e} 2009); 
(b) contamination by unrelated
absorption features due to lower redshift systems; and (c) limited
S/N ratio, makes it
very difficult in our view to deduce reliable ion column densities
from the analysis of the line profiles. Accordingly, we have refrained
from attempting to do so.

\subsection{The Lyman $\alpha$ line}
\label{sec:Lya}

\begin{figure}
\vspace*{-1.075cm} 
\centerline{\hspace{0.205cm}
  \includegraphics[width=0.775\columnwidth,clip,angle=270]{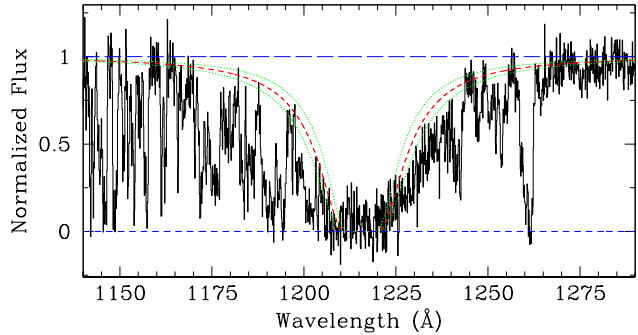}}
\vspace{-0.75cm}
\caption{Portion of the ESI spectrum of the lensed galaxy
in the Cosmic Eye encompassing the region of the
\lya\ line. Overlaid on the spectrum are 
three theoretical damped profiles, centred at
$z_{\rm sys} = 3.07331$, produced by column densities 
of neutral hydrogen 
$N({\rm H\,\textsc{i}}) = (3.0 \pm 0.8) \times 10^{21}$\,cm$^{-2}$.
}
\label{fig:DLA}
\end{figure}

The \lya\ line in the Cosmic Eye shows a strong damped
absorption profile, from which we deduce a neutral hydrogen
column density  
$N({\rm H\,\textsc{i}}) = (3.0 \pm 0.8) \times 10^{21}$\,cm$^{-2}$
(see Figure~\ref{fig:DLA}).
This is a high value of $N$(H\,{\sc i}), near the upper
end of the column density distribution of QSO-DLAs
(Noterdaeme et al. 2009) and more typical of the DLAs 
seen in the spectra of the afterglows of high redshift gamma-ray bursts
(Pontzen et al. 2010).
The corresponding line width is
${\rm FWHM} = 27.4$\,\AA\ or $\sim 6750$\,\kms\ (Jenkins 1971).
On kinematic grounds it is therefore impossible, with the S/N ratio of the present data,
to assign the DLA absorption to one or other of the two main components
visible in the metal absorption lines, which are separated by
only $\approx 400$\,\kms (Section~\ref{sec:kinematics}).

Nevertheless, we suspect that most of the DLA 
absorption may be due to the redshifted component,  which has
the larger covering fraction of the two
and consists primarily of neutral gas.  
The $\sim 70$\% covering
factor we surmised for the outflowing, blueshifted, component
may lead to the escape of a substantial fraction
of the \lya\ photons
produced by the H\,{\sc ii} regions surrounding the OB stars,
as found by Quider et al. (2009) in the case of the Cosmic Horseshoe.
And yet no \lya\ emission line is seen here, not even as residual
emission when the damped profile is divided out, as is the case
in cB58 (Pettini et al. 2000, 2002) and in 
the lensed $z = 3.77$ LBG FORJ0332$-$3557 observed by
Cabanac et al. (2008).
It is possible that any \lya\ photons that escape
from the star-forming regions in the Cosmic Eye are
absorbed by the foreground neutral gas giving rise to
the redshifted component
of the metal absorption lines.

It is interesting that the damped \lya\ line has a larger covering
factor than any other absorption feature in the spectrum.
Recall that at the outset we subtracted off 5\% of the continuum flux 
from the entire spectrum  (Section~\ref{sec:observations}), in order to
bring the core of the damped \lya\ line to zero.
At this small level, we cannot be sure whether this residual
flux in the core of the line is due to inaccurate background 
estimation, or is the true level of unobscured continuum flux.
In any case, the correction we have applied to our data
translates to a lower limit of 95\% to the covering factor
of the OB star continuum by foreground H\,{\sc i} gas.
The fact that this is larger than the values we deduced
in Section~\ref{sec:partial_cover} for the metal lines is not surprising,
given that for solar abundances the optical depth in the \lya\ line will
exceed that of even the strongest metal lines from H\,{\sc i} regions
by factors of $ > 10^4$. In other words, low column densities of
gas, undetectable even in the strongest metal absorption lines,
could be adding to the apparent optical depth in \lya\ to give
a higher overall covering fraction.

Bringing together the points discussed in this Section, 
our high resolution ESI observations have revealed that
the interstellar spectrum of the Cosmic Eye is unusual
in showing two absorption components, of approximately equal
strengths, one blueshifted by $-70$\,\kms\ and the other
redshifted by $+350$\,\kms\ relative to the systemic
redshift $z_{\rm sys} = 3.07331$ defined by the OB stars
and their H\,{\sc ii} regions. The redshifted component
is mostly of low ionization. The interpretation of the 
interstellar spectrum is complicated by the fact that the
metal lines appear to only partially cover the starburst,
with differing covering factors for blueshifted and redshifted
absorption.
There is a very large column of foreground neutral gas,
$N$(H\,{\sc i})\,$ = 3 \times 10^{21}$\,cm$^{-2}$,
giving rise to a strong damped \lya\ line which covers
at least 95\% of the stellar continuum.

\section{The Stellar Spectrum}
\label{sec:stars}

\begin{figure*}
\vspace{-1.75cm} \centerline{
  \includegraphics[width=1.65\columnwidth,clip,angle=270]{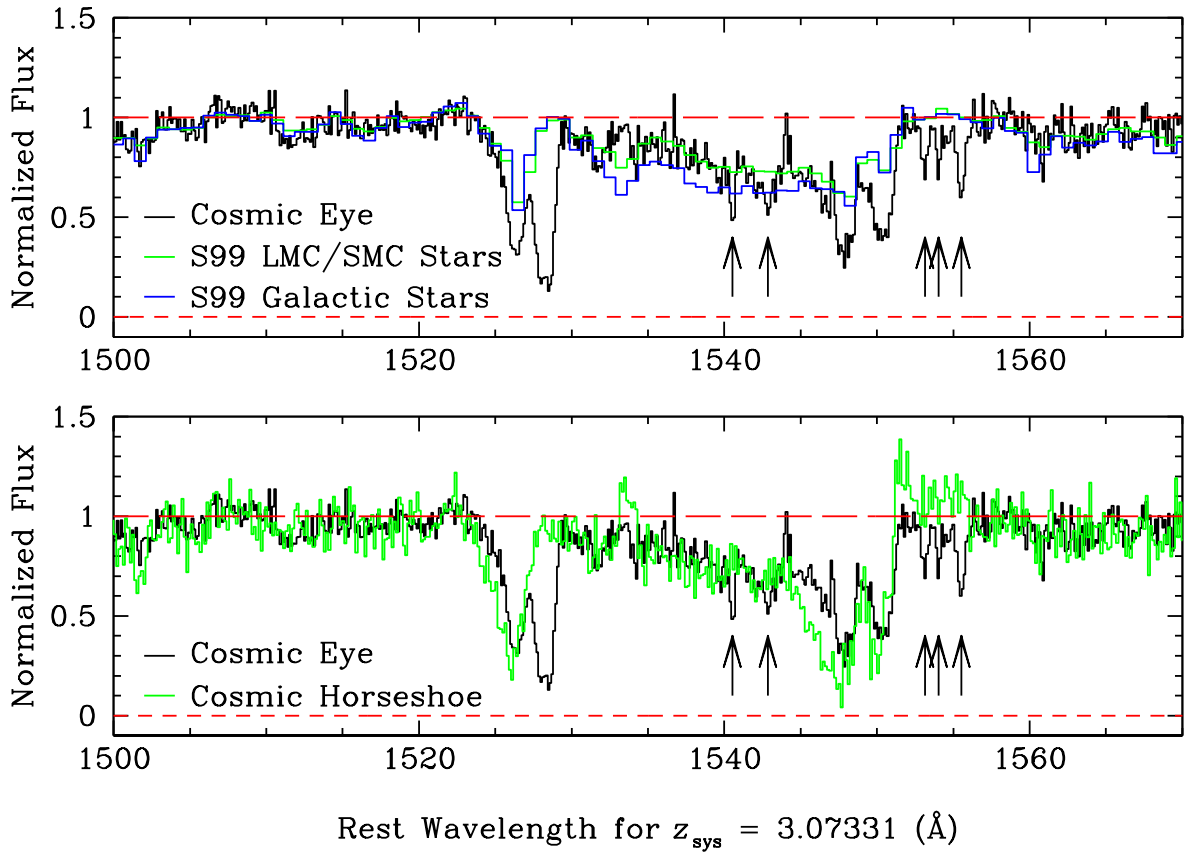}}
\vspace{-0.5cm}
\caption{\textit{Upper panel:} Comparison between the ESI spectrum of
  the Cosmic Eye in the region encompassing the C\,{\sc iv} line and
  two model spectra computed with \textit{Starburst99} and empirical
  libraries of, respectively,  Magellanic Clouds and Galactic stars, as indicated.
  The model spectra were generated assuming a 100\,Myr old continuous
  star formation episode with a Salpeter IMF with an
  upper mass limit $M_{\rm up} = 100 M_\odot$.  \textit{Lower panel:}
  The stellar spectra of the Cosmic Eye and of the Cosmic Horseshoe
  are remarkably similar in the wavelength region shown. The ESI
  spectrum of the Cosmic Horseshoe is reproduced from Quider et
  al. (2009), and has been reduced to its rest wavelengths at $z_{\rm
  stars} = 2.38115$.  
  Upward pointing arrows indicate the positions of narrow, intervening
  absorption lines unrelated to the Cosmic Eye itself 
  (see Section~\ref{sec:intervening}): a possible C\,{\sc iv} doublet 
  and three closely spaced Al\,{\sc ii}\,$\lambda 1670$ lines.
}
\label{fig:CIV}
\end{figure*}

The stars which provide the rest-frame UV continuum light in a high
redshift galaxy leave their imprint on the galaxy's spectrum.  The
stellar wind features are a clear signature from the most massive O
stars which have the strongest stellar winds (Kudritzki \& Puls 2000).
Rix et al. (2004) have shown that the P~Cygni profile of the stellar
wind features is sensitive to the metallicity of the stars,
as well as to the slope and
upper mass cut-off of the IMF.  While
stellar and interstellar features are usually blended at the low
resolution of typical high redshift galaxy spectra (Crowther et
al. 2006), the high resolution of our spectrum allows us to easily
separate them and focus on the stellar component.

We used model spectra computed with the spectral synthesis code
\textit{Starburst99} to exploit the metallicity and IMF
sensitivities of the P~Cygni profiles evident in the spectrum of the
Cosmic Eye.  The two model spectra considered here
were generated using
libraries of empirical UV spectra from stars either within the Galaxy
($Z \sim 1.0$\,\Zsun), or within the Large and Small Magellanic Clouds
($Z \sim 0.4$\,\Zsun; Leitherer et al. 1999, 2001).  Both models are
for 100\,Myr old continuous star formation with a Salpeter IMF
with upper mass limit $M_{\rm up} = 100 M_\odot$,
which are reasonable parameters for the current star formation episode (Coppin
et al. 2007; Pettini et al. 2000; Quider et al. 2009).

The strongest stellar wind feature present in the Cosmic Eye is the
C\,{\sc iv}\,$\lambda \lambda 1548, 1550$ P~Cygni profile shown in
Figure~\ref{fig:CIV}.  Also plotted in the top panel are the two
\textit{Starburst99} model spectra for comparison.  Clearly there is a
good agreement between the models and the data.  The main
deviations from the models are due to \emph{interstellar} 
Si\,{\sc  ii}\,$\lambda 1526$ and C\,{\sc iv}\,$\lambda \lambda 1548, 1550$
(see Section \ref{sec:ISM}), which are evident in the blue absorption
trough, and narrower intervening Al\,{\sc ii}\,$\lambda 1670$ and 
C\,{\sc  iv}\,$\lambda \lambda 1548, 1550$ absorption (see Section
\ref{sec:intervening}) which can be seen in the P~Cygni emission and
absorption respectively (the intervening absorptions 
are highlighted with upward pointing arrows in Figure~\ref{fig:CIV}).
Of the two model spectra shown, the one computed
with libraries of Magellanic Cloud stars is the better match,
while Milky Way metallicities overpredict the strength
of the absorption. 
This conclusion is also consistent with the P~Cygni profile
of N\,{\sc v}\,$\lambda \lambda 1238, 1242$, although the 
analysis of the latter is complicated by blending with the red wing
of the damped \lya\ line and some intervening absorption.
Thus, we infer that $Z_{\rm O\,stars} \approx 0.4\,Z_\odot$.  
We also investigated the effects on the emergent
P~Cygni profile of varying the slope
and upper mass cut-off of the IMF. 
As found in previous studies, 
(e.g. Pettini et al. 2000),
there are no indications of a departure from
the `standard'  Salpeter IMF for stars
with masses $M > 5 M_\odot$.

The lower panel of Figure \ref{fig:CIV} compares the P~Cygni region
of the Cosmic Eye with that of the Cosmic Horseshoe, also recorded with ESI
(reproduced from Quider et al. 2009).  The P~Cygni
profiles of these two galaxies are very similar, and both are
similar to that seen in cB58 (Quider et al. 2009).
The most striking difference between the Cosmic Horseshoe and
the Cosmic Eye is in their interstellar absorption profiles: the Horseshoe's
interstellar absorption extends to higher outflow velocities, 
while in the Cosmic Eye 
there is strong absorption 
at positive velocities in Si\,{\sc ii}\,$\lambda 1526$
(and other ions dominant in neutral gas),
as discussed above (Section~\ref{sec:kinematics}).
The emission component of the C\,{\sc iv} P~Cygni profile,
which can be clearly seen
in the Cosmic Horseshoe and cB58, seems to be absent 
from the Cosmic Eye.
Smail et al. (2007) interpreted this as evidence that
the massive stellar population of the galaxy is deficient
in O-type stars; with our higher spectral resolution 
we can now see that the lack of emission is partly due to 
the presence of foreground absorption by an intervening
triplet of Al\,{\sc ii}\,$\lambda 1670$ lines at
$z_{\rm abs} = 2.7865, 2.7891, 2.7925$ (see Table~\ref{tab:Intervening}).
Finally, nebular Si\,{\sc ii}\,$\lambda 1533.4312$ emission, 
which is evident in the Horseshoe and cB58, is weak in the Cosmic Eye.

Several blends of stellar photospheric
absorption lines in the integrated spectra of star-forming
galaxies have been shown to be sensitive to the 
metallicity of early-type stars (Leitherer et al. 2001; Rix et al. 2004).  
Two of these regions  are a blend of 
Si\,{\sc iii}\,$\lambda 1417$, C\,{\sc iii}\,$\lambda 1427$, and 
Fe\,{\sc v}\,$\lambda 1430$ absorption lines spanning the
wavelength interval 1415--1435\,\AA\ (the ``1425'' index 
of Rix et al. 2004), and a blend of several Fe\,{\sc iii} absorption lines 
from B stars between 1935 and 2020\,\AA\ 
(the ``1978'' index).
Previous work has shown the success of using the 1425 index
to estimate the stellar metallicity (Rix et al. 2004; Halliday et
al. 2008; Quider et al. 2009), but these same studies are divided on
the effectiveness of the 1978 index.  

In the case
of the Cosmic Eye, we cannot use the 1425 index because 
it is heavily contaminated by intervening absorption lines
at lower redshifts (Section~\ref{sec:intervening}). 
While these features are significantly narrower than the
broad blend of stellar absorption, they nevertheless
compromise the placement of the continuum in this region
of the spectrum, rendering the measurement of the 1425 index too
uncertain to be useful.
The 1978 index region appears to be `clean', but we found 
it to be a generally poor match to the model spectra by
Rix et al. (2004), in particular being stronger than any of
the metallicities considered by those authors between 
1985 and 2015\,\AA.
Quider et al. (2009) also
found the 1978 index to be problematic for the Cosmic Horseshoe
(although in that case the index failed between 1960 and 1980\,\AA), 
so we reiterate their cautionary note that the 1978 index needs to be
tested on more galaxies to establish its usefulness as an abundance
indicator.

To sum up this section, from a consideration of the C\,{\sc iv} stellar
wind profile, the population of early-type stars in the
lensed galaxy of the Cosmic Eye was found to be very similar to 
those of the Cosmic Horseshoe and cB58. In all three cases,
\textit{Starburst99} models with
metallicity $Z \simeq 0.4$\,\Zsun, continuous star
formation, and Salpeter IMF 
in the mass range $100 M_\odot \geq M \geq 5 M_\odot$
provide satisfactory fits to the UV spectra.
The 1978 index is
not well-matched by the spectrum of the Cosmic Eye at any of the model
metallicities, indicating that further testing of this index is
necessary before it can be reliably used as a metallicity measure.

\section{Intervening Absorption Systems}
\label{sec:intervening}


\begin{table*}
 \begin{minipage}{170mm}
 \caption{\textsc{Intervening Absorption Line Systems}}
 \begin{tabular}{llcc l}
\hline  \hline $\lambda_{\rm obs}$ (\AA)  & Identification & $z_{\rm
  abs}$ & $W_{\rm obs}$ (\AA) & Comments\\  \hline
\multicolumn{5}{c}{System 1: $z_{\rm abs} = 2.4563$}\\ \hline 
5277.15 & Si\,{\sc ii} 1526.7070 &
2.4567 & 1.31      & Blended with Si\,{\sc iv} 1393.7602 in System 4 and stellar absorption features
\\ 5351.27 & C\,{\sc iv} 1548.204 & 2.4564 & 1.80 &
\\ 5360.11 & C\,{\sc iv} 1550.781 & 2.4564 & 1.26 & Blended with Ni\,{\sc ii} 1317 in Cosmic Eye
\\ 5559.42 & Fe\,{\sc ii} 1608.4511 & 2.4563 & 0.44 &
\\ 5774.69 & Al\,{\sc ii} 1670.7886 & 2.4563 & 1.88   &
\\ 6409.68 & Al\,{\sc iii} 1854.7184 & 2.4559 & 0.69   & \\ 6438.52 &
Al\,{\sc iii} 1862.7910 & 2.4564 & 0.48   & \\ 8102.21 & Fe\,{\sc ii}
2344.2139 & 2.4563 & 0.85   & \\ 8235.86 & Fe\,{\sc ii} 2382.7652 &
2.4564 & 1.37  & \\

\hline \multicolumn{5}{c}{System 2: $z_{\rm abs} = 2.6597$}\\ \hline
5100.64 & Si\,{\sc iv} 1393.7602 & 2.6596 & 1.08 & \\  \ldots & Si\,{\sc
  iv} 1402.7729 & \ldots & \ldots & Blended with Si\,{\sc ii} 1260 in Cosmic
Eye\\ 5665.89 & C\,{\sc iv} 1548.204 & 2.6597 &1.07   &\\ \ldots &
C\,{\sc iv} 1550.781 & \ldots & \ldots & Blended with Si\,{\sc iv} 1393 in
Cosmic Eye and C\,{\sc iv} 1548 in System 3\\

\hline \multicolumn{5}{c}{System 3: $z_{\rm abs} = 2.6639$}\\ \hline
5106.67& Si\,{\sc iv} 1393.7602 & 2.6640 & 2.61  & Blended with
S\,{\sc ii} 1253 in Cosmic Eye\\ \ldots & Si\,{\sc iv} 1402.7729
& \ldots & \ldots & Blended with Si\,{\sc ii} 1260 in Cosmic Eye\\ 5672.94&
C\,{\sc iv} 1548.204 & 2.6642& 5.42  & Blended with Si\,{\sc iv} 1393
in Cosmic Eye and C\,{\sc iv} 1550 in System 2\\ 5682.19& C\,{\sc iv}
1550.781 &2.6641 &3.64  & Blended with Si\,{\sc iv} 1393 in Cosmic
Eye\\ 6794.61& Al\,{\sc iii} 1854.7184 &2.6634 &1.06   & Blended with
Al\,{\sc ii} 1670 in Cosmic Eye\\ 

\hline \multicolumn{5}{c}{System 4: $z_{\rm abs} = 2.7865$}\\ 
\hline 
5053.73 & C\,{\sc ii} 1334.5323 & 2.7869 &0.67 & \\
5277.15 & Si\,{\sc iv} 1393.7602 &
2.7864 & 1.31   & Blended with Si\,{\sc ii} 1526 in System 1 and stellar absorption features
\\ \ldots & Si\,{\sc iv} 1402.7729 & \ldots  & \ldots & Blended with
O\,{\sc i} 1302 in Cosmic Eye\\ 5780.88& Si\,{\sc ii} 1526.7070 &
2.7865 & 0.69  & \\ 5862.35& C\,{\sc iv} 1548.204 & 2 .7865 & 1.93   &
\\ 5871.72& C\,{\sc iv} 1550.781 & 2.7863 & 2.12   & Blended with
C\,{\sc iv} 1548 in System 6\\ 6090.54& Fe\,{\sc ii} 1608.4511 &
2.7866 & 0.78  & \\ 6326.42& Al\,{\sc ii} 1670.7886 & 2.7865 & 0.57
& Blended with C\,{\sc iv} P~Cygni emission in Cosmic Eye\\ 7023.16&
Al\,{\sc iii} 1854.7184 & 2.7866  & 0.60   & \\ 7053.51& Al\,{\sc iii}
1862.7910 & 2.7865 & 0.34   & \\

\hline \multicolumn{5}{c}{System 5: $z_{\rm abs} = 2.7891$}\\ 
\hline
5057: & C\,{\sc ii} 1334.5323 & 2.79: & \ldots & Blended \\ 
5281.97& Si\,{\sc iv} 1393.7602 & 2.7896 & 0.61  & Blended with stellar absorption features
\\ \ldots & Si\,{\sc iv} 1402.7729 & \ldots & \ldots & Blended with
Si\,{\sc ii} 1304 in Cosmic Eye\\ 5784.63& Si\,{\sc ii} 1526.7070 &
2.7890 & 0.83 & \\ 5866.87& C\,{\sc iv} 1548.204 & 2.7895 & 1.45   &
\\ 5876.32& C\,{\sc iv} 1550.781 & 2.7893 &1.00   & \\ 6330.05&
Al\,{\sc ii} 1670.7886 & 2.7887 & 0.41   & Blended with C\,{\sc iv} 
P~Cygni emission in Cosmic Eye\\ 7026.91& Al\,{\sc iii} 1854.7184 &
2.7887 &0.24    & \\ 7058.23& Al\,{\sc iii} 1862.7910 & 2.7891 &0.30
& \\ 
\hline \multicolumn{5}{c}{System 6: $z_{\rm abs} = 2.7925$}\\ 
\hline 
5061.13 & C\,{\sc ii} 1334.5323 & 2.7924 & 1.91 &\\
5789.76& Si\,{\sc ii} 1526.7070 & 2.7923 & 0.84
& \\ 5871.72& C\,{\sc iv} 1548.204 & 2.7926  & 2.12   & Blended with
C\,{\sc iv} 1550 in System 4\\ 5881.45& C\,{\sc iv} 1550.781 & 2.7926
&0.86   & \\ 6336.16& Al\,{\sc ii} 1670.7886 & 2.7923 &1.06    &
Blended with C\,{\sc iv} P~Cygni emission in Cosmic Eye\\ \hline
\multicolumn{5}{c}{System 7: $z_{\rm abs} = 2.8106$}\\ \hline 5899.50&
C\,{\sc iv} 1548.204 & 2.8105 & 1.49   & \\ 5909.52& C\,{\sc iv}
1550.781 & 2.8107 &1.25   & \\ \hline \multicolumn{5}{c}{(Possible) System 8:
  $z_{\rm abs} = 3.0528$} \\ \hline
6275.06& C\,{\sc iv} 1548.204 &3.0531 & 0.60   & Blended with C\,{\sc
  iv} P~Cygni absorption in Cosmic Eye\\ 6284.55& C\,{\sc iv} 1550.781
&3.0525  &0.58   & Blended with C\,{\sc iv} P~Cygni absorption in
Cosmic Eye\\ \hline
     \label{tab:Intervening}
\end{tabular}
 
 \end{minipage}
 \end{table*}

Possibly because of its high redshift, and the high resolution and S/N ratio
of our data, the spectrum of the Cosmic Eye shows numerous 
narrow absorption lines at redshifts $z < z_{\rm sys}$ due to
intervening absorbers along the line of sight.
Limiting ourselves to the wavelength interval longwards 
of the \lya\ line (so as to avoid the \lya\ forest), we
identified 41 such absorption lines associated
with seven definite and one possible systems 
at redshifts from $z_{\rm abs} = 2.4563$ to
$z_{\rm abs} = 3.0528$.  
Every system has a
C\,{\sc iv}\,$\lambda \lambda 1548,1550$ doublet associated with it and 
six systems have additional low or high ionization species present.
Table~\ref{tab:Intervening} lists the identified intervening
absorption lines with their redshifts and observed equivalent widths.

Although for clarity we have labelled each resolved set of absorption
lines as an `absorption system', some of the absorbers are separated by
redshift differences which correspond to relative velocities 
of only a few hundred km~s$^{-1}$. 
If, for ease of comparison with earlier
absorption line statistics from lower resolution work, 
we group together into one
`absorption system' lines which fall within a velocity interval
$\Delta v = 1000$\,km~s$^{-1}$, we have 
four such (definite) systems within a redshift interval $\Delta z = 0.81$
between $z_{\rm abs} = 2.25 $ and $z_{\rm abs} = 3.06$,
considering that we could have
detected C\,{\sc iv} doublets between $\lambda_{\rm obs} = 5030$\,\AA\
and $\lambda_{\rm obs} = 6293$\,\AA\ (approximately 
the redshifted wavelengths of \lya\ and C\,{\sc iv} in the Cosmic Eye).
All four systems thus defined have 
$W_0(1548) \geq 0.40$\,\AA, where $W_0(1548)$
is the rest-frame equivalent width of
the stronger member of the C\,{\sc iv} doublet.
Thus, in the Cosmic Eye the number of such systems per unit redshift
is $N(z) = 4/0.81 \simeq 5$, or $\sim 5$ times higher than
the mean $\langle N(z) \rangle \simeq 1$ for absorbers of this strength
and in the same redshift interval in QSO spectra (Steidel 1990). 
It is hard to conclude, without a more detailed
analysis which is beyond the scope of this paper, whether this is
just a statistical fluctuation, or whether the excess number of strong
C\,{\sc iv} absorbers in front of the Cosmic Eye is at least partly
due to the extension of the image on the plane of the sky (two arcs,
each $\sim 3^{\prime\prime}$ long). 

As can be seen in Figure \ref{fig:ISMvelocity3}, the intervening 
absorption lines are scattered
throughout the spectrum of the Cosmic Eye, inconveniently falling
within several features of interest. Column (5) of Table
\ref{tab:Intervening} details the blending of the intervening
absorbers with interstellar lines in the Cosmic Eye
as well as each other.  There is no
doubt that the presence of so many intervening absorption lines
complicates the interpretation of the Cosmic Eye spectrum.  This is
a clear case where a high resolution spectrum reveals a more complex
situation than was inferred from lower resolution observations.
Thus, the Cosmic Eye serves as a cautionary tale 
for the interpretation of high redshift galaxy spectra which,
without the boost provided by gravitational lensing, generally
can only be recorded at low resolution with available instrumentation. 

We also searched for Mg\,{\sc ii}\,$\lambda\lambda 2796, 2803$
absorption near $z = 0.73$, the redshift of the lensing galaxy of the Cosmic Eye
and of at least two other galaxies in the field (Smail et al. 2007).
However, these lines would fall in the \lya\ forest and, if present,
are difficult to disentangle from the intergalactic medium absorption
at the S/N ratio of our data.

\section{Discussion}
\label{sec:discussion}

One of the advantages offered by strongly lensed LBGs is that they
can be observed in numerous wavebands allowing us to put together 
a more extensive picture of their properties than is normally possible
in the absence of lensing. In this respect, 
the Cosmic Eye is among the better studied
galaxies at $z = 3$ (see Section~\ref{sec:eyesummary});
in this section we discuss our findings from the analysis of its
rest-frame UV spectrum in the light of previous work carried out 
at other wavelengths.

\subsection{Geometry and Reddening}

When Siana et al. (2009) 
compared the far-infrared (at wavelengths $\lambda = 40$--120\,$\mu$m, $L_{\rm FIR}$) 
and UV (1600\,\AA, $L_{\rm 1600}$) 
luminosities of the Cosmic Eye,
they found the ratio $L_{\rm FIR}/L_{\rm 1600}$
to be $\sim 8$ times lower than the value predicted with
the relationship by Meurer, Heckman, \& Calzetti  (1999), 
which relates $L_{\rm FIR}/L_{\rm 1600}$ to the slope $\beta$ of the 
UV stellar continuum in local starburst galaxies
(assumed to be a power law of the form $F_\lambda \propto \lambda^{\beta}$).
An analogous discrepancy, albeit by a smaller factor,
was found by Siana et al. (2008) for MS\,1512-cB58.
These two studies attributed such departures to a steeper
extinction curve for the dust in the Cosmic Eye and cB58
than that applicable to dust in local starbursts (Calzetti et al. 2000):
a faster rise in the extinction $A_\lambda$ with decreasing
$\lambda$ would have the net effect of 
increasing $\beta$ (i.e. making the UV spectral slope redder)
for a lower overall degree of dust extinction, as measured by the 
$L_{\rm FIR}/L_{\rm 1600}$ ratio. Siana et al. (2009) showed that
a UV extinction curve similar to those measured in the Large and 
Small Magellanic Clouds could account for the observed values
of $L_{\rm FIR}/L_{\rm 1600}$ and $\beta$ in the 
Cosmic Eye and cB58.

The difference between the Calzetti  and LMC/SMC extinction
curves is thought to be one of geometry, depending on 
whether the bulk of the dust is mixed with, or lies in front of,
the OB stars whose light is being attenuated. 
The implication that in cB58 and the Cosmic Eye most of the 
dust may be located in a `foreground screen' 
is consistent with the finding that in both cases 
the UV spectra of the galaxies exhibit a damped \lya\ line
with near 100\% covering factor (Pettini et al. 2002 for cB58,
and Section~\ref{sec:Lya} of the present paper for the Cosmic Eye).
The ubiquitous presence of galaxy-scale outflows in star-forming
galaxies at $z = 2 - 3$ may lead to a geometrical configuration
of dust and stars more analogous to that described by the 
LMC/SMC curves in many LBGs, as considered by Pettini et al. (1998a).
Another factor may be age: in their recent study, Reddy et al. (2010)
find that it is the galaxies that are younger than $\sim 100$\,Myr that
exhibit a tendency  
to lie below the Meurer et al. (1999) relation.

Looking ahead, 
with a larger sample of high resolution spectra of strongly lensed LBGs
it may be possible to disentangle the effects of geometry and age
on the emergent UV flux of star-forming galaxies.
An immediately obvious test is to measure the 
$L_{\rm FIR}/L_{\rm 1600}$ ratio in LBGs where the \lya\ line
is predominantly in emission, such as the Cosmic Horseshoe
where our earlier work also showed that the 
insterstellar medium only covers
 $\sim 60 \% $ of the UV stellar continuum (Quider et al. 2009).  
 In the physical picture put forward by Shapley et al. (2003)
 and more recently Kornei et al. (2009),
 the three parameters (age, \lya\ luminosity, and covering factor)
 are related, in that galaxies
 with strong \lya\ emission represent a later evolutionary stage,
 in which supernova-induced outflows have reduced the dust covering 
 fraction.

\subsection{Kinematics}
\label{sec:discuss_kinematics}
The interstellar absorption lines provide information on the
kinematics of the interstellar gas along our line of sight to the OB
stars of the Cosmic Eye.  This only allows us to make a
one dimensional assessment of the gas motions in the galaxy,
given that the UV light is dominated by a single source 
with a half-light radius of $\sim 1$\,kpc
in the lensing model of Dye et al. (2007---see Section~\ref{sec:eyesummary}).
However, using laser-guided adaptive optics, Stark et al. (2008)
were able to probe the two-dimensional distribution of velocities
of nebular emission lines
([O\,{\sc iii}] and H$\beta$) across the face of the galaxy 
on unprecedented small scales of $\sim 100$\,pc.
They found a regular pattern (see Figure~2 of Stark et al. 2008)
from which they constructed a rotation curve extending 
to $\pm 2$\,kpc from the dynamical centre with an amplitude
of $v_{\rm rot} \sin i  = 55 \pm 7$\,\kms, where $i$ is the
inclination angle. Superposed on this regular pattern are
chaotic motions with a dispersion $\sigma_0 = 54 \pm 4$\,\kms.
Such a low ratio of ordered to random motions,
$v_{\rm rot}/\sigma_0 \sim 1$, is commonly found
in star-forming galaxies at $z = 2$--3 
(e.g. Law et al. 2009; F\"{o}rster Schreiber et al. 2009).\footnote{
These estimates refers to the inner regions ($\sim 2$\,kpc)
of the galaxy; larger motions may well be found if one were
able to probe baryons over the full virial radius of the galaxy.}
 
To these kinematic data we now add measurements of 
outflowing gas seen in absorption. 
As discussed in Section~\ref{sec:kinematics},
the blueshifted component of the absorption lines
has maximum apparent optical depth at a velocity
$v_{\rm blue} \simeq -70$\,\kms, relative to $z_{\rm sys} = 3.07331$,
and extends out to $v_{\rm blue}^{\rm max} \simeq -500$\,\kms\
(see Figure~\ref{fig:ISMvelocity3}).
In their recent analysis of outflows in low redshift
starburst galaxies,  Martin \& Bouch\'{e} (2009)
associate the velocity of maximum apparent optical depth
with the speed of a shell of swept-up, interstellar gas 
at the time of blowout from the disk of the galaxy,
at a few pressure scale heights. In their picture,
gas at higher negative velocities is
located further away from the disk (and is therefore
accelerating), and exhibits lower apparent optical depths
as a result of geometrical dilution.
If this scenario also applies to the Cosmic Eye,
we would conclude that rotation, velocity dispersion,
and outflow speed of swept-up interstellar matter 
at blowout are all of comparable magnitude.
However, it is still far from clear how
all these motions fit into one coherent
picture.

It is interesting that the outflow speeds in the Cosmic Eye
are lower than the values measured in most other LBGs.
The value $v_{\rm blue} \simeq -70$\,\kms\ is only about 
half the mean blueshift of the strongest interstellar
lines in the composite spectrum of 811 LBGs constructed
by Shapley et al. (2003),  
and lower than $v_{\rm blue} \simeq -150$\,\kms\
and $v_{\rm blue} \simeq -255$\,\kms\
measured in the Cosmic Horseshoe (Quider et al. 2009)
and MS\,1512-cB58 (Pettini et al. 2002) respectively.
The maximum blueshift at which
absorption is detected,  $v_{\rm blue}^{\rm max} \simeq -500$\,\kms\
is also lower than $v_{\rm blue}^{\rm max} \simeq -750$\,\kms\
in both the Horseshoe and cB58. 
However, this limit may simply be an observational one,
determined by the decrease in covering factor
with distance from the central starburst, 
and there may be gas at higher negative velocities which is too
diluted to produce detectable absorption (Martin \& Bouch\'{e} 2009). 
While there are indications at both low ($z \sim 0$, Martin 2005; 
Rupke et al. 2005) and intermediate 
($z \simeq 1.4$, Weiner et al. 2009) redshifts
that `superwind' speeds scale with galactic mass
and star formation rate,
there is sufficient scatter in these relations to accommodate 
the differences we have uncovered between the three
strongly lensed high-$z$ galaxies, even though they are thought to have
comparable masses and star formation rates.

Finally we note that, alongside rotation, random motions,
and outflows, our UV spectrum has shown 
the presence of high column density, neutral
gas apparently moving \textit{towards} the OB stars in the Cosmic Eye,
further complicating the kinematic picture of this galaxy.
With the available data, we cannot arrive at definite conclusions 
concerning the nature and location of this gas. 
One possibility is that it may be a chance superposition of another
galaxy (or damped \lya\ system) along the line of sight.
Alternatively, it may be dynamically related
to the Cosmic Eye. For example, it may be gas
ejected from the galaxy by a previous episode of 
star formation and now falling back onto it, or we may be
viewing a merger along a favourable
sightline, although in both cases the 350\,\kms\ velocity 
offset from $z_{\rm sys}$ seems very high, given that the Cosmic
Eye does not appear to be a very massive galaxy
[e.g. $M_{\rm stars+gas} \leq 1 \times 10^{10} M_\odot$ (Coppin et al 2007),
and $v_{\rm rot} \sin i = 55$\,km~s$^{-1}$ (Stark et al. 2008)].
The lensing model of Dye et al. (2007) does include
a second, separate clump of UV light in the source plane, which 
could indicate a merger, or a foreground galaxy.
However, Coppin et al. (2007) reported 
a redshift $z_{\rm CO} = 3.0740 \pm 0.0002$
for the CO emission which they found to be peaked 
at the position of this
second source; the small velocity difference $\Delta v = 50 \pm 20$\,\kms\
from the systemic redshft $z_{\rm sys} = 3.07331 \pm 0.00024$ 
(Section~\ref{sec:z_sys}) makes it unlikely that the absorbing gas 
we have found with $v_{\rm red} \simeq +350$\,\kms\ is associated
with the CO emitting clump.

Additional kinematic data on strongly lensed galaxies spanning 
a range of physical properties are clearly needed
in order to explore how outflows (and in some cases presumably 
inflows) tie into the other motions
within a galaxy, as well as the galaxy's intrinsic characteristics.

\subsection{Metallicity and Stellar Populations}

The C\,{\sc iv} P~Cygni profile encodes information about the
metallicity of the O stars, as well as the relative numbers
of massive stars in the Cosmic Eye. Here we consider each in turn.

Up to now, the degree of metal enrichment attained
by high redshift star-forming galaxies 
has been determined primarily from the analysis of the most
prominent nebular lines emitted
from their H\,{\sc ii} regions (e.g. Pettini et al. 2001; 
Shapley et al. 2004; Erb et al. 2006a; Maiolino et al. 2008).  
Strongly lensed galaxies give us the opportunity 
to conduct a complementary metallicity measurement 
based on UV spectral features due to massive stars. 
Young stars and H\,{\sc ii} regions are expected
to have intrinsically the same chemical composition,
since the former have only recently formed out of the gas
which they ionize into the latter; thus a comparison between
stellar and nebular abundances is essentially a consistency
check which may reveal systematic offsets between different
metallicity indicators.
In the case of the Cosmic Eye, 
Stark et al. (2008) deduced 
a metallicity $Z_{\rm H\,\textsc{ii}} \simeq 0.9 Z_\odot$
assuming that the upper-branch solution of the R23 method
first introduced by Pagel et al. (1979) applies.
On the other hand, we showed in Section~\ref{sec:stars}
that at solar metallicities the C\,{\sc iv} P Cygni profile
would be stronger than observed, and that a metallicity
$Z_{\rm LMC/SMC} \approx 0.4 Z_\odot$ gives a 
better match to the data. 
Discrepancies by factors of $\sim 2$ between different 
metallicity indicators in high redshifts galaxies are
probably to be expected (Pettini 2006).
On the other hand, in our earlier study of the 
Cosmic Horseshoe (Quider et al. 2009), 
we did find the metallicity
deduced from the R23 index to be the odd one out
(in the sense of being a factor of $\sim 3$ higher)
among several stellar and nebular measures.
At lower redshifts too, methods relying on the R23
index result in metallicities  towards the upper  
end of the range spanned by different indicators
(see, for example, Figure 2 of Kewley \& Ellison 2008).
In light of these considerations, the apparent difference
between nebular and stellar metallicity
in the Cosmic Eye is not surprising.

As mentioned in Section \ref{sec:stars}, the C\,{\sc iv} P~Cygni
profiles of the Cosmic Eye, Cosmic Horseshoe, and MS\,1512-cB58 are 
all remarkably similar.  
While initially this may seem to be an odd coincidence, considering 
that these three galaxies were randomly selected by gravitational 
lensing, on further reflection this result is perhaps to be expected.
The key issue here is continuous star formation.
When considering the composite spectrum of an entire
galaxy,\footnote{The UV luminosity of an $L^{\ast}$
$z = 3$ LBG is equivalent to the integrated output
of $\sim 2.5 \times 10^5$ O7 stars (Pettini et al. 1998b).}
it is exceedingly unlikely that we would pick out a special
time in its stellar evolution. While in individual regions
star formation is likely to proceed in bursts,
a succession of such bursts on a galaxy-wide scale
will approximate a continuous process
of star formation. Under such circumstances,
the contrast of the C\,{\sc iv} (and other) wind
lines over the underlying OB photospheric continuum
will stabilize after $\sim 50$\,Myr from the onset
of star formation, corresponding to the lifetime
of the lowest mass stars contributing to the light
at 1550\,\AA. Since this time interval is comparable
to the dynamical timescale of most LBGs (e.g. Erb et al. 2006c),
only a fairly exceptional  star-formation
history would produce significant changes in the integrated
UV stellar spectrum.

As discussed above, the stellar metallicity is also an important
factor in determining the strengths of the P~Cygni lines,
but on this basis too we would not expect marked differences
 between the Cosmic Eye, the Cosmic Horseshoe
and MS\,1512-cB58, given that the three galaxies 
have similar metallicity,   $Z \simeq 0.4 Z_\odot$.
Such a uniform degree of metal enrichment is in turn not as 
surprising as it may seem, since the three objects are
of comparable luminosities, reflecting the fact that most strongly
lensed galaxies targeted for detailed analyses so far
are intrinsically luminous, 
with $L \sim L^\ast$.  
A test of these ideas would be provided
by observations of highly magnified galaxies
of lower intrinsic luminosity (and presumably metallicity)
than the few examples studied up to now.

\section{Summary and Conclusions}
\label{sec:conclusions}

Strong gravitational lensing of a $z = 3.07331$ star-forming 
[${\rm SFR} \simeq 50\,M_\odot$~yr$^{-1}$ for a Chabrier (2003) IMF] 
galaxy magnifies it by a factor of $\sim 25$ and distorts its
image into two $3^{\prime\prime}$ long arcs
which have been collectively named the `Cosmic Eye'
by their discoverers, Smail et al. (2007).
The high
magnification has allowed us to use the ESI spectrograph
on the Keck\,{\sc ii} telescope to record the galaxy's rest-frame 
UV spectrum with high resolution and S/N ratio. By analyzing these data 
together with existing observations of the Cosmic Eye
at other wavelengths, we have reached the following main conclusions.

(i) The interstellar absorption lines exhibit two components,
of approximately equal strength, which are respectively
blueshifted by  $-70$\,\kms\ and redshifted by $+350$\,\kms\ 
relative to the stars and H\,{\sc ii} regions.  
While these values apply to the gas with the highest apparent optical depths,
both components include absorption spanning  several hundred \kms.
We associate the blueshifted component 
with a galaxy-wide outflow similar to, but possibly weaker than,
those seen in most star-forming galaxies at $z = 2$--3.
The redshifted absorption is very unusual, and may represent 
gas ejected by a previous episode of star formation and now
falling back onto the galaxy, or a merger viewed along a favourable 
line of sight. Alternatively, it may just be a chance superposition
of another galaxy along the line of sight.

(ii) Both components of the metal absorption lines
show indications that they do not
fully cover the OB stars against which they are being viewed;
we estimate covering fractions of $\sim 70$\% and 
$\sim 85$\%  for the blueshifted and redshifted component
respectively. There must also be more pervasive diffuse
gas, because the strong damped \lya\ line, corresponding
to a column density $N$(H\,{\sc i})\,$=  (3.0 \pm 0.8) \times 10^{21}$\,cm$^{-2}$,
covers at least 95\% of the UV stellar continuum.
We tentatively associate this high column density of H\,{\sc i} 
with the redshifted component of the metal lines,
where absorption from ionized species is weak or missing altogether,
and propose that it provides the `foreground screen' of dust
responsible for the lower-than-expected far-infrared luminosity
of the Cosmic Eye found 
by Siana et al. (2009)
with the \textit{Spitzer Space Telescope}.

(iii) The internal kinematics of the galaxy lensed into the Cosmic Eye
are very complex, our data now adding outflow,
and possibly even inflow,
to the rotation and velocity dispersion already known from the
integral field
spectroscopy with adaptive optics by Stark et al. (2008).  
Ordered rotation, chaotic motions, and outflow all seem to
be of comparable magnitude, with 
$v_{\rm rot} \sin i \approx \sigma_0 \approx v_{\rm blue} \simeq  50$--70\,\kms.
However, we do not have a model yet of how these different
motions fit together into one coherent kinematic picture.

(iv) Turning to the stellar spectrum, we find that
the C\,{\sc iv} P~Cygni profile is well fit by a \textit{Starburst99}
stellar population model spectrum having continous star formation
with a Salpeter IMF, stellar masses from 5 to 100\,\msun, and a
LMC/SMC metallicity of $Z \sim 0.4$\,\Zsun.  
The P~Cygni profiles of
the Cosmic Eye, the Cosmic Horseshoe, and MS\,1512-cB58, three
high redshift star-forming galaxies studied at high spectral resolution, 
are all nearly identical. 
This is not unexpected, however, when we consider
that in each case we see the integrated light
of several hundred thousand O-type stars,
 and that these three galaxies have
similar metallicities and dynamical timescales over which star
formation is taking place.

(v) The metallicity $Z \simeq 0.4 Z_\odot$ deduced for the 
O stars in the Cosmic Eye is lower by a factor of $\sim 2$ 
than the
value derived from the analysis of the strong nebular
lines from its H\,{\sc ii} regions using the R23 index.
We consider this apparent discrepancy to reflect systematic
offsets between different abundance estimators, rather
than intrinsic inhomogeneities in the chemical composition
of stars and ionized gas.

(vi) The interpretation of both interstellar and stellar
features in the UV spectrum of the Cosmic Eye is complicated
by the presence of numerous intervening absorption lines
associated with eight absorption systems at redshifts
$z_{\rm abs} = 2.4563$--3.0528.  
These narrow features are not resolved in existing low-resolution data, 
highlighting the caution that should be exercised in interpreting the
spectra that are typically available for unlensed LBGs.

In closing, the new data presented here further emphasize
the complexity of the physical conditions which prevailed in
actively star-forming galaxies at redshifts $z = 2$--3.
It is remarkable that the three strongly-lensed galaxies 
targeted by our ESI observations, while showing very similar
young stellar populations, are all different in the detailed
properties of their interstellar media. Whether such
variety is simply the result of different geometries 
and viewing angles, or has its roots in more fundamental
physical reasons remains to be established. Fortunately, 
with the increasing attention being given to detailed studies
of gravitationally lensed galaxies, we can look forward 
with optimism to a more comprehensive empirical picture  
of galaxy formation coming together in the years ahead.

\section*{Acknowledgements}
We are grateful to the staff at the W.~M. Keck Observatory for their
competent assistance with the observations. 
Valuable suggestions by an anonymous referee improved
the final version of this paper.
AMQ's research is funded by
a scholarship from the Marshall Foundation,
and a National Science Foundation graduate research fellowship.  
AES acknowledges support
from the David and Lucile Packard Foundation and the Alfred P. Sloan
Foundation, and CCS from NSF grant AST-0606912 and the John D. and
Catherine T. MacArthur Foundation.  
DPS's research is support by the UK
STFC through the award of a postdoctoral fellowship.
Finally, we wish to extend thanks
to those of  Hawaiian ancestry on whose mountain we are privileged to
be guests.

\label{lastpage}

\end{document}